# Parametric amplification of random lattice soliton swinging

Yaroslav V. Kartashov, Victor A. Vysloukh, and Lluis Torner

*ICFO-Institut de Ciencies Fotoniques, Mediterranean Technology Park, and Universitat Politecnica de Catalunya, 08860 Castelldefels (Barcelona), Spain*

We consider the propagation of solitons in media with an imprinted transverse periodic or parabolic refractive index modulation when the modulation depth slightly fluctuates along the propagation direction. We find that, under suitable spectral resonance conditions, small transverse soliton oscillations may get parametrically amplified.

The possibility to control the evolution of light beams in nonlinear materials is important from fundamental and practical viewpoints. Materials with shallow transverse refractive index modulations afford a unique exploration laboratory [1,2]. Longitudinal modulation of transversally periodic guiding structures allows managing their diffraction properties [3,4] and enables diffractionless propagation [5]; periodically curved waveguide arrays allow self-imaging of linear patterns [6,7] and support diffraction-managed solitons [8]; dynamical lattices can drag solitons [9,10], while longitudinal modulation in waveguides results in periodic modes conversion [11]. Importantly, a variety of parametric phenomena can occur in guiding structures with an imprinted harmonic longitudinal modulation [12-14]. However, exact harmonic longitudinal modulation may not be desired in practice, so that an important question arises about whether parametric phenomena are possible in randomly modulated refractive index landscapes.

In this Letter we show that parametric amplification of soliton transverse oscillations, or "swinging", is possible in periodic and parabolic refractive index landscapes in the presence of shallow random longitudinal refractive index variations. The amplification efficiency is given by the central frequency and by the spectral width of the random modulation.

Our analysis is based on the nonlinear Schrödinger equation describing the propagation of an optical beam inside a slab nonlinear waveguide, where the beam diffracts along the $x$ axis, while the refractive index is modulated along both axes. Namely,

$$i\frac{\partial q}{\partial \xi} = -\frac{1}{2}\frac{\partial^2 q}{\partial \eta^2} - q|q|^2 - pQ(\xi)R(\eta)q. \qquad (1)$$

Here $q(\eta,\xi) = (L_{\text{dif}}/L_{\text{nl}})^{1/2} A(\eta,\xi) I_0^{-1/2}$, $A(\eta,\xi)$ is the slowly varying field amplitude, $I_0$ is the input peak intensity, $\eta = x/r_0$, $r_0$ is the beam width, $\xi = z/L_{\text{dif}}$, $L_{\text{dif}} = r_0^2 n_0 \omega/c$ is the diffraction length, $L_{\text{nl}} = c/\omega n_2 I_0$ is the nonlinear length, $\omega$ is the carrying frequency, $p = L_{\text{dif}}/L_{\text{ref}}$ is the guiding parameter, $L_{\text{ref}} = c/(\delta n \omega)$ is the linear refraction length, $\delta n$ is the modulation depth. We consider harmonic modulation $R(\eta) = \cos(\Omega_l \eta)$, where $\Omega_l$ is spatial frequency, and parabolic modulation with $R(\eta) = 1 - (\Omega_l \eta)^2/2$. Note, that in such materials as AlGaAs ($n_2 \simeq 10^{-13}$ cm$^2$/W) at $\lambda = 1.53\,\mu$m and $r_0 = 5\,\mu$m diffraction length amounts to $L_{\text{dif}} \simeq 0.3$ mm, while typical nonlinearity-induced refractive index change $\delta n \sim 0.8 \times 10^{-3}$. The characteristic linear refractive index modulation depth in the lattice that we consider is $\sim 0.2 \times 10^{-3}$ and lattice period $\sim 30\,\mu$m.

The variation of the refractive index along the $\xi$-axis is described by a random function $Q(\xi) = 1 + \mu N(\xi)$, where $\mu < 1$ is the modulation depth, and $N(\xi)$ is the stochastic process with Gaussian power spectral density $P(\Omega) = (2\pi S^2)^{-1/2} \exp[-(\Omega - \Omega_n)^2/S^2]$ with the central frequency $\Omega_n$ and width $S$. In the narrow-band limit $N(\xi)$ transforms into the harmonic function $N(\xi) = \cos(\Omega_n \xi + \varphi_0)$ where the random phase $\varphi_0$ is uniformly distributed at $[-\pi, \pi]$. Narrow-band stochastic process with $S \ll \Omega_0$, $\Omega_n \sim \Omega_0$ (see definition of $\Omega_0$ below) is referred as "colored noise", while the broad-band process with $S > \Omega_0$, $\Omega_n = 0$ is termed "white noise".

An effective particle approach yields equation $d^2\eta_c/d\xi^2 + \Omega_0^2[1 + \mu N(\xi)]\eta_c - \gamma \eta_c^3 = 0$ describing dynamics of the integral center $\eta_c(\xi) = U^{-1}\int_{-\infty}^{\infty} |q|^2 \eta d\eta$ of soliton solution $|q(\eta,\xi=0)| = \text{sech}(\eta)$ of Eq. (1) obtained at $p = 0$. Here $U = \int_{-\infty}^{\infty} |q|^2 d\eta$. This equation is analogous to that describing a nonlinear pendulum parametrically driven by a *random* force. Here $\Omega_0^2 = p\Omega_l^2(\pi\Omega_l/2)\sinh^{-1}(\pi\Omega_l/2)$ defines the frequency of free oscillations at $\mu = 0$, while $\gamma = \Omega_0^2 \Omega_l^2/6$ is the nonlinearity parameter that accounts for anharmonicity of oscillations. In the case of small-amplitude oscillations $\Omega_l \eta_c \ll 1$ one gets Mathieu-type equation for $\eta_c$, while in parabolic potential $\gamma \equiv 0$. Further we assume that light is launched at $\eta = 0$, so that $\eta_c(0) = 0$ and $(d\eta_c / d\xi)_{\xi=0} = \alpha_0$, where $\alpha_0$ is the incident angle. In the limit $S \to 0$ and for $\gamma \to 0$ the trivial solution $\eta_c = 0$ is unstable when the resonance condition $\Omega_n = 2\Omega_0 / (1+m)$ (here $m = 0,1,2...$) is satisfied. Thus, small perturbations of $\eta_c$ grow exponentially with the increment $\delta \sim \mu\Omega_n$. The width of the principal instability band $\Delta\Omega$ centered at $\Omega_n = 2\Omega_0$ is then given by $\Delta\Omega = \mu\Omega_0$. In the case of parametric amplification of seed oscillations $\eta_c(\xi) = A(\xi)\sin(\Omega_0\xi)$, where $A\big|_{\xi=0} = \alpha_0 / \Omega_0$, $\delta = (\mu\Omega_0 / 4)\sin\varphi_0$ one finds that the increment of small oscillations growth is given by . The point is, this growth depends critically on the random phase shift $\varphi_0$. In particular, if $\varphi_0 \simeq \pi/2$ oscillations are exponentially amplified, while at $\varphi_0 \simeq -\pi/2$ they fade away. Note that when $\gamma \neq 0$ the maximal amplitude $A_{max} = 2\Omega_0(\mu/\gamma)^{1/2}$ of oscillations is limited by $\gamma$.

To elucidate dynamics of soliton swinging we used the Monte-Carlo approach and integrated Eq. (1) numerically with a split-step Fourier method up to a fixed distance $L$ for different computer-generated noise series $N_m(\xi)$, $m = 1,...,M$. We calculated the random trajectories of the integral soliton center $\eta_m(\xi)$, propagation angles $\alpha_m(\xi) = d\eta_m / d\xi$, as well as the accumulated kinetic energies $E_m(\xi) = \alpha_0^{-2}\int_0^\xi \alpha_m^2(\xi)d\xi$. The statistical averaging of the latter energy $K(\xi) = M^{-1}\sum_{m=1}^M E_m(\xi)$ provides information about the mean growth rate of the soliton swinging. We fixed parameters: $\Omega_l = 1$, $p = 0.2$, $\alpha_0 = 0.02$, and $L = 246$. Note that for such values the frequency of free oscillations amounts to $\Omega_0 = 0.461$. We varied the central noise frequency $\Omega_n$ and its width $S$.

Figure 1 shows propagation of solitons in a parabolic potential for two samples of colored noise with $\Omega_n = 2\Omega_0$. Since the noise spectral bandwidth is rather small ($S = 0.0153$, which is around 3.3% of $\Omega_0$) the swinging amplification is mostly determined by $\varphi_0$. Thus, Fig. 1(a) shows dynamics of swinging at $\varphi_0 \simeq -\pi/2$ that causes initial fading of oscillations, while Fig. 1(b) shows amplification of oscillations at $\varphi_0 \simeq \pi/2$. Figure 1(c) illustrates growth of $E$ with distance in a logarithmic scale for several samples of colored noise having larger bandwidth $S = 0.051$. The random shift of these curves is related to the initial random phases $\varphi_0$, while changes in curve slopes are determined mostly by the random variations of $N(\xi)$ amplitude. As expected, the amplitude of the random soliton swinging depends crucially on $\Omega_n$. The resonance curve of the accumulated kinetic energy is shown in Fig. 2(a). The maximal value of $K(L)$ is achieved at the frequency of the principal para-metric resonance $\Omega_n = 2\Omega_0$, and it decreases monotonically with the growth of $S$ [Fig. 2(b)]. In harmonic lattices the resonant frequency is smaller [Fig. 2(c)] because of the "soft" nonlinearity of the restoring force. In such lattices $K(L)$ decreases with $S$ slower than in parabolic potentials [Fig. 2(d)] and typical values of the accumulated kinetic energies are remarkably smaller since with increasing amplitude, the frequency of free oscillations drops off and the system detunes from resonance.

The Monte Carlo approach allows estimation of the probability density functions. Figure 3(a) shows histograms of random accumulated kinetic energy in parabolic potentials ($\Omega_n = 2\Omega_0$, $S = 0.0077$). A feature of this histogram is the appearance of two local maxima of the probability density. Such twin-peak profile might be qualitatively explained taking into account the fact that at $S \ll \Omega_0$ the fluctuations of $\varphi_0$ play a major role; the increment is given by $\delta = (\mu\Omega_0/4)\sin\varphi_0$, and that the accumulated kinetic energy can be expressed as $E(L) = [\exp(2\delta L) - 1]/2\delta$ [see Fig. 3(c)]. Notice that with growth of $S$ a maximum of the histogram at high $E(L)$ ceases to exist [Fig. 3(b)] because the correlation length $L_{\text{cor}} = 2/S \ll L$ and thus the probability of unfavorable random phase and amplitude changes upon propagation substantially increases. Dynamics of random swinging under the influence of white noise ($\Omega_n = 0$) is completely different (Fig. 4). If such noise has narrow bandwidth, only weak low-frequency parametric resonances $\Omega_n = 2\Omega_0/(1+m)$ with $m \gg 1$ are involved and the growth of $K(L)$ is almost negligible. However, when $S$ increases the noise spectrum overlaps with the strongest principal resonance ($\Omega_n = 2\Omega_0$), and $K(L)$ increases and reaches its maximum [Fig. 4(c)]. A subsequent growth of $S$ causes a decrease in $K(L)$ because the power spectral density within the resonance curve contour again drops off. Figure 4(a) shows histograms of the accumulated kinetic energy at $\mu = 0.2$. If longitudinal modulation depth increases [Fig. 4(b)] the probability density distribution spreads up and the most probable value of $E$ grows, so the observed behavior may be viewed as a thermalization of soliton swinging.

Summarizing, we showed that parametric amplification of soliton swinging is possible in periodic and parabolic potentials in the presence of random shallow longitudinal modulations of the refractive index. The efficiency of swinging depends critically on the spectral width of the random longitudinal modulation and on the overlap of such spectrum with the resonant curve corresponding to the strongest parametric resonance.

# References with titles

# References without titles

# Figure captions

Figure 1. Random soliton swinging in parabolic potential for two different samples of colored noise at $S = 0.0153$. In (a) random phase is close to $\varphi_0 \simeq -\pi/2$, while in (b) it is close to $\varphi_0 \simeq \pi/2$. (c) The growth of accumulated kinetic energy versus propagation distance for different realizations at $S = 0.0510$. In all cases $\Omega_n = 2\Omega_0$, $\mu = 0.2$, $\alpha_0 = 0.02$, and $p = 0.2$.

Figure 2. Accumulated kinetic energy: (a) in parabolic potential, versus $\Omega_n$ at $S = 0.0153$, (b) in parabolic potential, versus $S$ at $\Omega_n = 2\Omega_0$, (c) in harmonic potential, versus $\Omega_n$ at $S = 0.0153$, (d) in harmonic potential, versus $S$ at $\Omega_n = 1.72\Omega_0$. In all cases $\mu = 0.2$, $\alpha_0 = 0.02$, $p = 0.2$, $L = 256$, $M = 300$.

Figure 3. Histograms of $E(L)$ for parabolic potential at (a) $S = 0.0077$ and (b) $S = 0.0153$. In both cases $\Omega_n = 2\Omega_0$, $p = 5.2$, $\mu = 0.2$, and $\alpha_0 = 0.02$. (c) The histogram for the case of monochromatic noise ($S \to 0$) with fluctuating initial phase $\varphi_0$ uniformly distributed on the segment $[-\pi, \pi]$.

Figure 4. Random swinging under the influence of broadband white noise at $\Omega_n = 0$, $p = 0.2$. Histograms of $E(L)$ in parabolic potential at (a) $\mu = 0.2$ and (b) $\mu = 0.25$. (c) Accumulated kinetic energy $K(L)$ versus white noise bandwidth $S$ for harmonic potential (curve 1) and for parabolic potential (curve 2) at $\mu = 0.2$, $M = 300$.

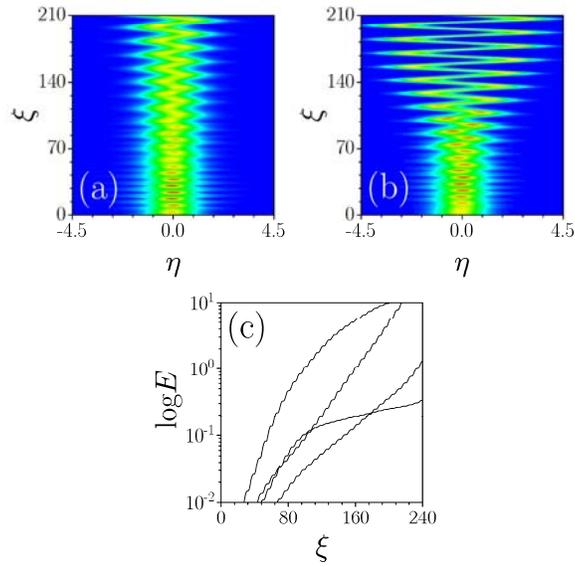

Figure 1. Random soliton swinging in parabolic potential for two different samples of colored noise at $S = 0.0153$. In (a) random phase is close to $\varphi_0 \simeq -\pi/2$, while in (b) it is close to $\varphi_0 \simeq \pi/2$. (c) The growth of accumulated kinetic energy versus propagation distance for different realizations at $S = 0.0510$. In all cases $\Omega_n = 2\Omega_0$, $\mu = 0.2$, $\alpha_0 = 0.02$, and $p = 0.2$.

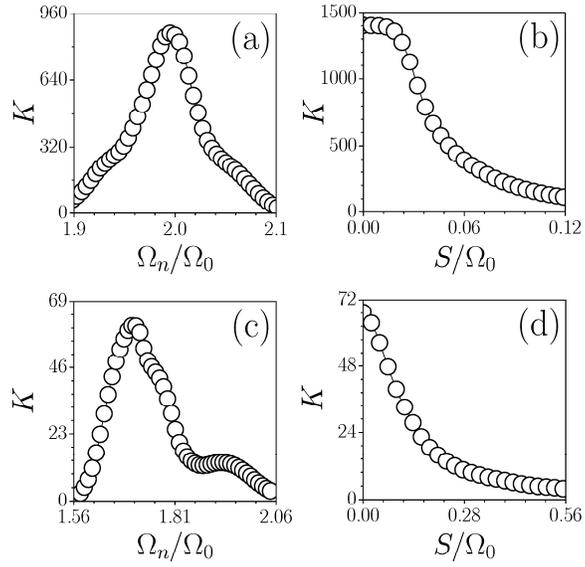

Figure 2. Accumulated kinetic energy: (a) in parabolic potential, versus $\Omega_n$ at $S = 0.0153$, (b) in parabolic potential, versus $S$ at $\Omega_n = 2\Omega_0$, (c) in harmonic potential, versus $\Omega_n$ at $S = 0.0153$, (d) in harmonic potential, versus $S$ at $\Omega_n = 1.72\Omega_0$. In all cases $\mu = 0.2$, $\alpha_0 = 0.02$, $p = 0.2$, $L = 256$, $M = 300$.

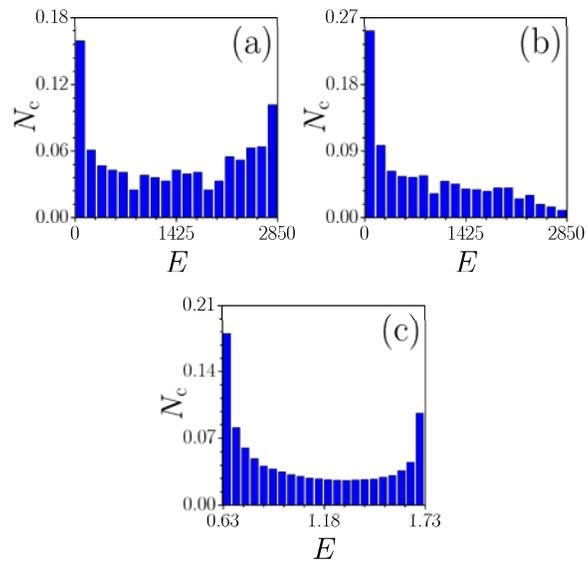

Figure 3. Histograms of $E(L)$ for parabolic potential at (a) $S = 0.0077$ and (b) $S = 0.0153$. In both cases $\Omega_n = 2\Omega_0$, $p = 5.2$, $\mu = 0.2$, and $\alpha_0 = 0.02$. (c) The histogram for the case of monochromatic noise ($S \to 0$) with fluctuating initial phase $\varphi_0$ uniformly distributed on the segment $[-\pi, \pi]$.

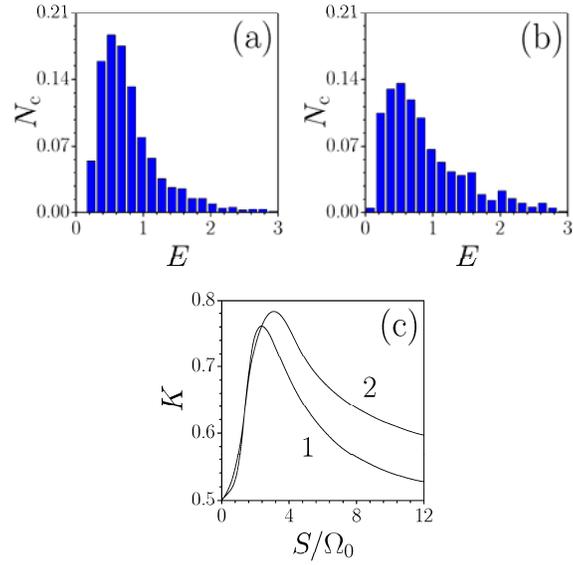

Figure 4. Random swinging under the influence of broadband white noise at $\Omega_n = 0$, $p = 0.2$. Histograms of $E(L)$ in parabolic potential at (a) $\mu = 0.2$ and (b) $\mu = 0.25$. (c) Accumulated kinetic energy $K(L)$ versus white noise bandwidth $S$ for harmonic potential (curve 1) and for parabolic potential (curve 2) at $\mu = 0.2$, $M = 300$.

12